\documentclass[12pt, epsf]{article}
\input epsf.tex
\def\be{\begin{equation}} \def\ee{\end{equation}}
\def\bi{\begin{itemize}} \def\ei{\end{itemize}}
\def\bea{\begin{eqnarray}} \def\eea{\end{eqnarray}} \def\ba{\begin{array}}
\def\ea{\end{array}} \def\ben{\begin{enumerate}} \def\een{\end{enumerate}}

\newcommand{\eqn}[1]{(\ref{#1})}

\newcommand{\prl}[3]{Phys. Rev. Lett. {\bf#1} ({#2}) {#3}}

\newcommand{\prd}[3]{Phys. Rev. {\bf D#1} ({#2}) {#3}}
\newcommand{\hepth}[1]{{\tt [arXiv:{#1}[hep-th]]}}

\def\br{\nonumber\\}

\begin{document}
{}~
\hfill \vbox{\hbox{arXiv:1309.nnnn} 
\hbox{\today}}\break

\vskip 3.5cm
\centerline{\Large \bf
 Schr\"odinger  Spacetimes with Screen}
\centerline{\large \bf
and }
\centerline{\large \bf
Reduced Entanglement}

\vskip 1cm

\vspace*{.5cm}

\centerline{\bf  Harvendra Singh}

\vspace*{.5cm}
\centerline{ \it  Theory Division} 
\centerline{ \it  Saha Institute of Nuclear Physics} 
\centerline{ \it  1/AF Bidhannagar, Kolkata 700064, India}
\vspace*{.25cm}

\vspace*{.5cm}

\vskip.5cm

\vskip1cm

\centerline{\bf Abstract} \bigskip

We study a particular class of  type II string  vacua   
which become Schr\"odinger like spacetime in the IR region but
are conformally AdS  in asymptotic UV region. These solutions are
found to possess some unique properties such as the presence of
a spacetime `screen'. This Schr\"odinger (spacetime) screen is however very 
different from a black-hole horizon. It 
requires the presence of  finite chemical potential and a negative charge
density in the Schr\"odinger CFT.  We find that these 
vacua give rise to  reduced entanglement entropy as 
compared to   Lifshitz-AdS 
counterpart, perhaps due to the screening effects. 

\vfill 
\eject

\baselineskip=16.2pt


\section{Introduction}

A steady progress \cite{son}-\cite{Faulkner:2013ana} has been made
towards understanding  the string solutions  which exhibit
Lifshitz and Schr\"odinger type nonrelativistic symmetries. 
Particularly, in these solutions 
the time and space coordinates in the dual CFT 
scale {\it asymmetrically}.
 Some of these systems exhibit a non-fermi liquid or strange metallic 
 behaviour at very low temperatures. 
These strange effects have been alluded to the 
fact that strongly correlated quantum 
systems might have hidden fermi surfaces \cite{takaya11, subir11}. 
There are similar  issues related to 
the entanglement of information in the  quantum systems \cite{RT, takaya11}. 
The entanglement of the subsystems is a very common concept in quantum physics
in general, including black-holes \cite{Maldacena:2013xja},
as well as an entangled quantum EPR pair \cite{Faulkner:2013ana}.
But when a strongly correlated 
system at  critical point can be represented as a
system living on the boundary of some  bulk gravity  theory, 
the subject becomes much more phenomenologically appealing. In such 
holographic cases  
the entanglement
entropy of a subsystem in the boundary can  be defined geometrically 
as the area of a minimal surface, lying within the bulk spacetime having
specific boundary conditions \cite{RT}.  

Following an early work on $AdS_5\times S^5$ 
and finding the Lifshitz solutions
\cite{hs10}, we recently  generalized that  approach for all
 D$p$-brane AdS vacua and obtained
Lifshitz and Schr\"odinger like solutions   
in type II string theory \cite{hs12}.\footnote{As per our terminology,
all the Lifshitz-like
solutions we discuss in this paper will have appropriate
 conformal factors in front of the Lifshitz metric 
when viewed after the compactification. 
Thus we shall be discussing `conformally Lifshitz' vacua all through. 
These give rise to scaling violations in the nonrelativistic CFT. 
The corresponding hyperscaling  parameters are calculated in \cite{hs12}. 
Thus our solutions are different from the Lifshitz solutions introduced in 
\cite{kachru}, or the solutions with light-like matter \cite{nara1}.
Similarly,
the Schr\"odinger solutions  also come with appropriate conformal factors.}   
   These vacua exhibit a fixed amount of supersymmetry.  
Our primary focus in this work are particularly 
the Schr\"odinger-like  D$p$ brane solutions \cite{hs12}. 
They appear in various dimensions  and have
nontrivial dynamical exponents given by $a_{_{\rm Sch}}={2\over p-5}$, 
which is negative for 
$p<5$. These IR Schr\"odinger solutions 
can be said to be the least understood type vacua, at least
if we ask the questions about the entanglement entropy of the boundary CFT. 
The special characteristic of these zero temperature
solutions is that they  involve a compact null direction, 
because of which one generally
cannot trust these Schr\"odinger geometries for classical calculations
in the bulk geometry. 
It is so  because for  a meaningful (nonrelativistic)
Schr\"odinger CFT description to arise, 
such as described in \cite{son,bala,malda}, it 
requires the null lightcone coordinate 
to be  compact.  In this work,
 we study a new class of Schr\"odinger-like 
 D$p$-brane solutions, which are closely related to the solutions
 given in \cite{hs12}, thus keeping the essential facts unchanged. 
However, the method we 
employ is applicable to
any other Schr\"odinger  solution. 
  We shall  engineer our solutions such that 
they first become well behaved classical geometries, at least 
in some finite UV region, 
such that they could eventually be compactified.
Our ultimate aim is  to
 estimate the entanglement entropy of the CFT living at the boundary of these
Schr\"odinger vacua.
The new solutions are such that they  interpolate  smoothly
between ``conformally
Schr\"odinger'' solutions in the IR and the ``conformally AdS'' spacetime 
in the UV. We demonstrate
various properties of these classical solutions in 
both the   regions. 
These interpolating  solutions are then used to calculate the
entanglement entropy of strip-like subsystem in the boundary CFT. 
In conclusion, we find that a Ryu-Takayangi entropy functional can be suitably
defined. We must emphasize here that  
for the standard Schr\"odinger metric such as $a=2$ \cite{son,bala} $$ 
-{(dx^{+})^2\over z^4}  +{-dx^{+}dx^{-}
+d\vec x^2\over z^2}
+ {dz^2\over  z^2} $$
the Ryu-Takayanagi entanglement entropy cannot be defined. 
In these Galilean $a=2$ vacua, $x^-$ is  a compact `null' coordinate,
and because of this we cannot find a spacelike extremal surface inside 
the bulk whose area would give the entanglement entropy. Particularly
all constant $x^+$ (time) surfaces have vanishing area. We will avoid this
situation altogether in this work.

The paper is planned as follows. In  section-2 we first review 
interpolating Lifshitz-AdS  
D$p$-brane vacua and some of the 
basic properties including the entanglement entropy. 
The expert reader  can skip this section 
and directly shift to section-3.  
In  the section-3, we construct  `conformally Schr\"odinger'  solutions which  
asymptotically become `conformally AdS' vacua. These solutions give rise to  
finite chemical potential
and a (negative) charge density. This construction allows us to
introduce a new concept  of  {\it Schr\"odinger spacetime screen}, 
for  these asymptotically AdS  solutions. We  define the
entanglement entropy  for the interpolating  solutions. It is found that
the entanglement entropy is lower for Schr\"odinger-AdS cases as compared to
 the Lifshitz-AdS case, provided the global parameters
are kept the same. 
Some numerical analysis is presented in section-4  
to reinforce our conclusions. 
 The summary is provided  in the section-5.

\section{ Interpolating Lifshitz-AdS string vacua }

We first review the interpolating  Lifshitz like solutions \cite{hs13} as 
these are close cousins of Schr\"odinger type solutions which we will be
discussing in the next section.
The Lifshitz  D$p$-brane  solutions with eight supersymmetries
are given as \cite{hs12}
\bea\label{sol3}
&&ds^2_{lif}=R_p^2 z^{p-3\over p-5} \bigg[ 
\{ {\beta^2\over z^{4/(p-5)}} (dx^{-})^2 +{-dx^{+}dx^{-}
+d\vec x_{(p-1)}^2\over z^2}  
+{4\over (5-p)^2} {dz^2\over  z^2} \}  + d\Omega_{(8-p)}^2 \bigg] ,\br
&& e^\phi=(2\pi)^{2-p}(g_{YM})^2 R_p^{3-p}{ z^{(7-p)(p-3)\over 2(p-5)}}
\eea 
with the $(p+2)$-form RR-flux (for $p\ne 5$). 
Here $\beta$ is an arbitrary  parameter
and it can also be absorbed by  scaling the lightcone coordinates. 
Note that the metric component
$g_{--}> 0$, so $x^-$ is a space-like coordinate while $x^+$ 
will be treated as the lightcone time.  
When $\beta=0$ 
these solutions exactly become  conformally AdS (for $p\ne 3$) string
solutions which arise as the near horizon geometry of $N$ coincident
 D$p$-branes. 

In  the solutions \eqn{sol3}  
the lightcone coordinates
scale  {\it asymmetrically} under the dilatations 
\be\label{sol4}
z\to \xi z , ~~~~
x^{-}\to \xi^{2-a}  x^{-},~~~~
x^{+}\to \xi^{a} x^{+},~~~\vec x\to \xi \vec x
\ee 
with the dynamical exponent  
 $a={2p-12\over p-5}$.  Note,  under this scaling
the string coupling, $e^\phi$, and 
the string metric  in eq.\eqn{sol3} do  get 
conformally rescaled (for $p\ne 3$). Thus for $p\ne 3$ the scaling \eqn{sol4}
is not a symmetry of the solutions. This  
is a well known RG property of the D$p$-brane AdS vacuas; see eq.
\eqn{d2} in the Appendix. Upon lightcone compactification,
various conformal factors in the metric \eqn{sol3}
contribute to the
overall hyperscaling (conformal scaling) property of the Lifshitz metric, 
 see next eq.\eqn{pok9a}.
 Note that $x^-$ spatial coordinate of the D$p$-brane 
has a different scaling as compared to
the remaining $(p-1)$ spatial coordinates $x^i$'s. 
The latter $p-1$ coordinates do  have 
a rotational symmetry amongst them. 
We are mainly interested in the situation where $x^-$ is taken
a compact coordinate. Note it is the compactification of $x^-$
 that gives us a "conformally Lifshitz"  metric  in
lower dimensions. Thus the solutions \eqn{sol3}
have all that  one would require for a hyperscaling 
Lifshitz spacetime description 
at the IR fixed point \cite{hs12}. 

As emphasized above
 we shall have to take $x^-$  being compact  so as to get 
 the actual Lifshitz like solutions. 
We write down explicit compactification of the  solutions  
\eqn{sol3}. 
 Upon  compactifications along the circle $x^{-}$, also along 
the sphere $S^{8-p}$, 
we get the  $(p+1)$-dimensional conformally  Lifshitz metric 
(given in the Einstein frame) 
\be\label{pok9a}
ds^2_{lif_{p+1}}\sim
z^{2(p^2-6p+7)\over (p-1)(p-5)}\left( -{(dx^{+})^2\over \beta^2 
z^{2 a_{lif}}}
+{d\vec{x}_{p-1}^2  +dz^2\over  z^2}\right) .
\ee
Note these 
Lifshitz  metrics have a specific conformal factor, of the type  
$z^{2\theta \over p-1}$, and the
 corresponding fact is summarized by defining a 
 hyperscaling parameter $\theta$ 
\be\label{sol6}
\theta_{lif}={p^2-6p+7\over p-5},
\ee
which  characterises these Lifshitz string solutions. 
Note that $\theta_{lif}$ is never vanishing 
for these conformally Lifshitz solutions \eqn{pok9a}. 
Actually these solutions essentially
describe a nonrelativistic Lifshitz dynamics at the
 IR fixed point. This description
 is however valid over a limited holographic range only. 
For example, these solutions cannot be good solutions for  
UV description of a Lifshitz  CFT, because in the UV region 
 the size of $x^-$ circle in the geometry would become sub-stringy. In 
other words the effective string coupling will diverge in the UV. 
Hence for these solutions to have 
a suitable UV description  we have had 
to modify them and
attach appropriate asymptotic AdS configuration, see \cite{hs13,hs10a}. 
That program leads us to the interpolating class of Lifshitz vacua.
\begin{figure}[htbp]
\begin{minipage}[t]{3in}
\centerline{\epsfxsize=2.5in
\epsffile{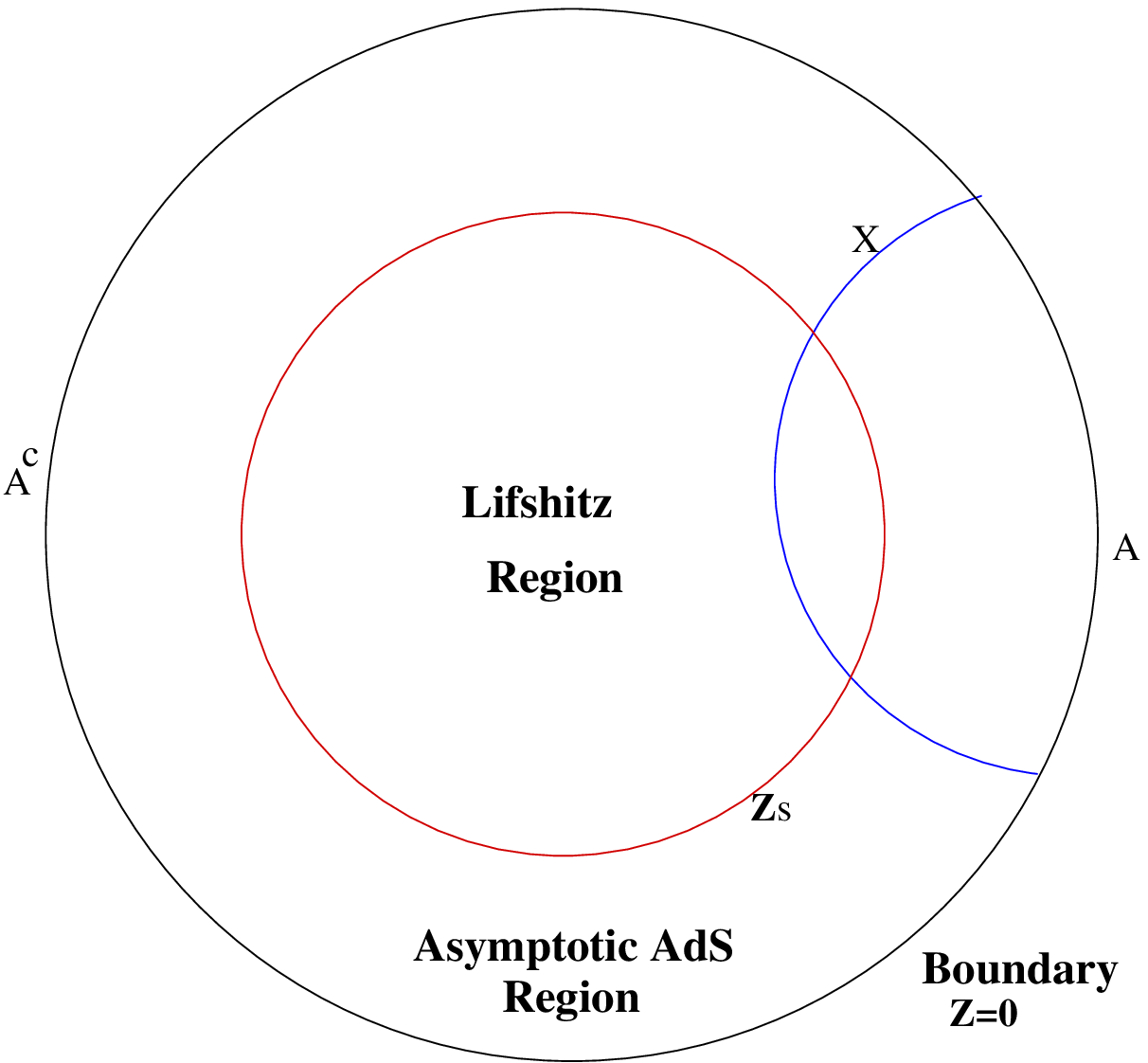} }
\caption{\label{figure3} \it The central Lifshitz
region  ends at
$z=z_{s}$ and smoothly connects to asymptotic AdS region. }
\end{minipage}
\hspace{0.4cm}
\begin{minipage}[t]{3in}
\centerline{\epsfxsize=2.5in
\epsffile{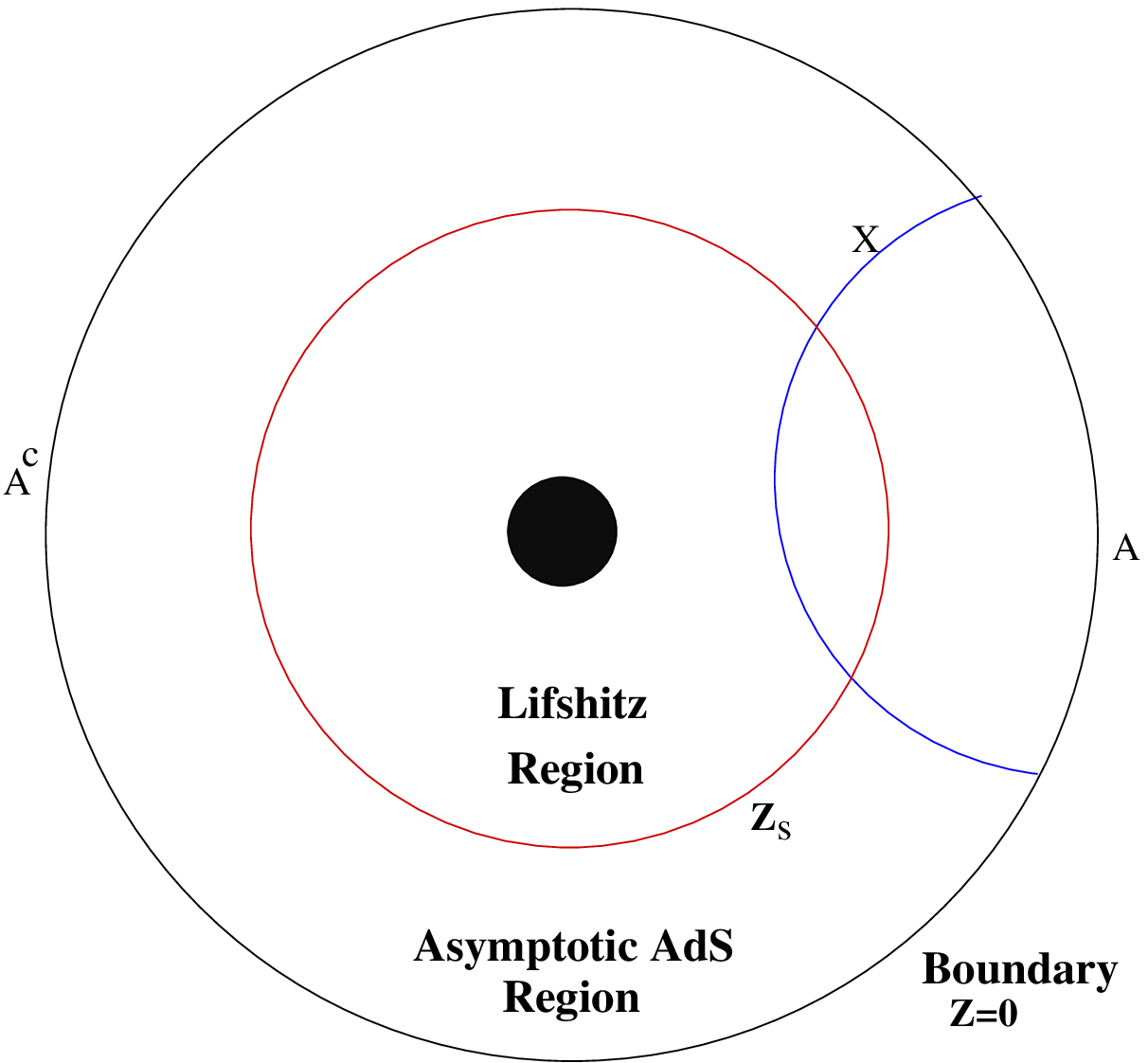} }
\caption{\label{figure3a} \it The Lifshitz-AdS
spacetime  with a  black hole at the center. 
Any minimal surface $X$ can reach only upto the horizon.}
\end{minipage}
\end{figure}

The interpolating  Lifshitz-AdS  solutions 
can be written as (for $p\ne 5$) \cite{hs13}
\bea\label{int03}
ds^2_{Lif-AdS}
&=&R_p^2 z^{p-3\over p-5} \bigg[ 
\{ {K\over z^2} (dx^{-})^2 +{-dx^{+}dx^{-}
+d\vec x_{(p-1)}^2\over z^2}  
+{4\over (5-p)^2} {dz^2\over  z^2}\}  + d\Omega_{(8-p)}^2 \bigg] \br
e^\phi&=&(2\pi)^{2-p}g_{YM}^2R_p^{3-p}{ z^{(7-p)(p-3)\over 2p-10}}
\eea 
with the same $(p+2)$-form flux. 
The new function 
\be\label{int032}
K(z)=v+ \left({z\over z_{IR}}\right)^{2p-14 \over p-5}
\equiv v\left(1+ \left({z\over z_{s}}\right)^{2p-14\over p-5}\right)
\ee 
is a harmonic function and plays the role of an interpolating function.
The parameter $z_{IR}>0$ is
an intermediate IR scale and is related to $\beta$ given earlier. 
Note that the solution \eqn{int03} is 
  interpolating solution because now the metric  \eqn{int03} 
smoothly connects Lifshitz and asymptotic AdS regions, provided $v$
is finite. Having finite $v$ does imply a presence of
a  chemical potential in the CFT \cite{hs13}.
\footnote{Note that in Ref.\cite{hs13} we simply took $v=1$.} 
In the asymptotic region, $z\ll z_{IR}$, and there $K\approx v$,
the solution \eqn{int03}
  starts behaving like  conformally AdS geometry ($a=1$). 
While in the IR region  $z\gg z_{IR}$,  where 
$K\approx ({z\over z_{IR}})^{2(p-7)\over p-5}$,
 it  behaves like a Lifshitz spacetime. 
Note that,
since these solutions are interpolating  configurations 
the scaling properties of the metric 
\eqn{int03} will not be explicit at the intermediate scales.
The  {\it  scaling}  property of
the metric  will become explicit only
in the neighbourhood of the  respective  IR  or  UV fixed points. 
An explicit lightcone 
compactification of the interpolating
 solutions \eqn{int03} provides following $(p+1)$-dimensional
Lifshitz spacetime (Einstein frame metric)
\bea\label{int035}
&&ds^2_{p+1}
=L^2 z^{2 \theta_{rel}\over (p-1) }K^{1\over p-1} \bigg[ 
 -{(dx^{+})^2\over 4 z^2 K}+ {d\vec x_{(p-1)}^2\over z^2}  
+{4\over (5-p)^2} {dz^2\over  z^2}   \bigg]   
\eea 
where $K$ is  given above in \eqn{int032}. Additionally
there is always a running $(p+1)$-dimensional dilaton field 
\bea
&& e^{-2\phi_{(p+1)}}\sim z^{p-5\over2} \sqrt{K} 
\eea and a Kaluza-Klein
 gauge field
\bea\label{hbn1}
&&A_{(1)}= -{1\over 2K}dx^{+},
\eea
and the flux component of RR-form. 
Note the $\theta_{rel}={p^2-7p+14 \over p-5}$ in the above 
is the effective hyperscaling parameter in the asymptotic UV region. 
The parameter $L$ is an  specific size factor which directly 
follows from compactification. One can see that
near $z\sim0$, $K$ becomes a constant and the metric becomes 
conformally AdS type, as it should be for the relativistic ($a=1$) CFT.
We must emphasize here, that these descriptions are valid so long as
string coupling remains perturbative and the spacetime curvature 
remains small. Beyond these UV scales, we must lift our solutions
to M-theory for valid holographic description, see for example \cite{itzhaki}.

Since these solutions interpolate between two asymptotia,  the value of
hyperscaling parameter $\theta$ will switch in between 
\be
\theta_{lif} \ge \theta\ge \theta_{rel}\ee
and correspondingly the
dynamical exponent will vary in between \be
a_{lif}\ge a \ge a_{rel}.\ee
For the conformally AdS spacetime  $a_{rel}=1$ and 
$\theta_{rel}<0$ \cite{hs13}.

Expanding the time component of the
gauge field \eqn{hbn1}
 in the neighborhood of the boundary ($z\sim 0$), we find (for $p=2,3,4$)
\bea
A_{+}(z\sim 0)= -{1\over 2 r^- v}
\left(1- {1\over v} 
\left( {z\over z_{IR}}\right)^{2p-14\over p-5}+\cdots \right).
\eea
 Thus 
 the chemical potential and the charge (number) 
density can be found to be 
\bea\label{kju1}
\mu=-{1\over 2 r^- v} , ~~~~~\rho={N\over V_{(p-1)}}
\simeq {(R_p)^{3p-6}{(r^-)^2}
  \over  G_{p+2}}  
\left({1\over z_{IR}}\right)^{2p-14\over p-5}
\eea
Here $G_{p+2}$ is the Newton's constant in $(p+2)$ dimensions, 
and $V_{(p-1)}$ is the regulated spatial
volume of the $p$-dimensional system.
Note we should never directly set $v=0$ in these expressions as these are
defined for finite $v$ only. Since the chemical potential is nonvanishing,
the CFT has a finite energy density  $E\sim \mu\cdot \rho$. We have 
dropped  overall normalisation  factors in the expressions in equation 
\eqn{kju1}. 

The figure \eqn{figure3}
represents a Lifshitz-AdS geometry with a minimal hypersurface $X$ suspended
inside the bulk geometry, while the figure \eqn{figure3a} depicts that
there could be a black hole inside Lifshitz geometry in the nonextremal case.
Using the Lifshitz metric \eqn{int035}
the entanglement entropy of the strip-like subsytem in the boundary   
can be obtained as \cite{hs13}
\bea\label{lifee3}
S_{Ent}={V_{p-2}L^d\over 2 G_{p+1}}
\int_{z_\ast}^{\infty}dz ~z^{9-p\over p-5} {2\over (5-p)} 
{ K
\over\sqrt{ v
+({z\over z_{IR}})^{2(p-7)\over p-5}
-C^2 z^{2(p-9)\over p-5}  }}
\eea  
where $C$ is an integration constant which 
depends on the turning point $z_\ast$
of the given extremal surface. The  extremality equation is 
\bea\label{akl3}
{dx_1\over dz}&=& {2\over 5-p} {C z^{p-9\over p-5}\over
\sqrt{v
+({z\over z_{IR}})^{2(p-7)\over p-5}
-C^2 z^{2(p-9)\over p-5}}} \br
\eea 
These are the  same expressions as were obtained in \cite{hs13} except that there
we took $v=1$. One should also refer to  the noncompact AdS-plane wave 
study in \cite{narayan12}. Our expressions naturally match with them. But we discuss
compactified situation and 
the parameter like $v$
has got definite interpretation in the compactified (Lifshitz) case, 
as giving rise to finite chemical potential in the boundary CFT. 
We avoid further discussion here, also a
very parallel calculation will appear in the next section.

\section{Schr\"odinger  vacua with a spacetime screen} 
The string solutions with Schr\"odinger symmetry group
 have been  well described  
in the initial works \cite{son, bala, malda}. Particularly
 Schr\"odinger  solutions with dynamical exponent $a=2$ 
could be constructed out of $AdS_5\times S^5$,
by employing 
a combination of null Melvin twist and pair of T-dualities, involving a fibered
direction along $S^5$ \cite{bala}. These correspond to the
 irrelevant operator deformations in the boundary CFT, so
those solutions  assume relativistic configurations in the IR. 
We are here interested in the IR 
 Schr\"odinger  solutions    
constructed
by taking the double limits of boosted `bubble AdS solutions' in
\cite{hs12}. 
These are given by \cite{hs12}
\bea\label{schsol3}
&&ds^2_{Sch}=R_p^2 z^{p-3\over p-5} \bigg[ 
\{ -{\beta^2\over z^{4/(p-5)}} (dx^{+})^2 +{-dx^{+}dx^{-}
+d\vec x_{(p-1)}^2\over z^2}  
+{4\over (5-p)^2} {dz^2\over  z^2} \}  + d\Omega_{(8-p)}^2 \bigg] ,\br
&& e^\phi=(2\pi)^{2-p}(g_{YM})^2 R_p^{3-p}{ z^{(7-p)(p-3)\over 2(p-5)}}
\eea 
with the $(p+2)$-form RR flux  ($p\ne 5$), being similar to the Lifshitz case. 
It can be noted that these Schr\"odinger vacua are  related via  
Wick rotations 
(along lightcone coordinates $x^+, x^-$) of the
corresponding Lifshitz solutions  \cite{hs12}. 
One significant difference between \eqn{sol3} and \eqn{schsol3} is that 
in the Lifshitz case it is the metric component $g_{--}$  
which is nontrivial and positive definite,
while in the Schr\"odinger solutions it is $g_{++}$
which is nontrivial and is a negative quantity. The other background 
fields such as the dilaton and the $(p+2)$-form flux however remain the
same in both type of nonrelativistic solutions. Actually
this relationship  is reminiscent of the known
fact that the black  D$p$-branes and 
the bubble D$p$-branes 
get  mapped into each other under  double Wick rotations too. 

In the above Schr\"odinger solutions  the lightcone coordinates
 scale  {\it asymmetrically} in a specific manner,  
under the dilatations 
$$
z\to \xi z , ~
x^{+}\to \xi^{2\over p-5}  x^{+},~
x^{-}\to \xi^{p-7\over p-5} x^{-},~\vec x\to \xi \vec x
$$ 
while the dilaton field and 
the metric in eq.\eqn{schsol3} conformally rescale as in eq.
\eqn{d2}. Thus the dynamical exponent of time in Schr\"odinger solutions
is \cite{hs12} $$a_{sch}={2\over p-5}$$
which is negative for $p<5$. We shall be mainly interested in these cases only.

It is however very crucial to bear in mind that 
 these Schr\"odinger  vacua cannot be immediately
compactified along the $x^-$  direction,
because $x^-$ is actually a null coordinate, i.e. $(g_{--}=0)$.
So the nonrelativistic interpretation of the boundary 
CFT remains the least understood concept for the
Schr\"odinger solutions, because such classical bulk geometry  \eqn{schsol3}
cannot be used for any useful calculations. 
Thus what should be our approach to
have a meaningful nonrelativistic Schr\"odinger description. 
A remedy was suggested in \cite{malda}
whereby one could include a black hole in the interior of the
asymptotically (UV) Schr\"odinger vacua. But this led to a finite temperature CFT.
Here instead we have got IR Schr\"odinger solutions. In this work  we 
shall try to  modify our IR solutions 
by adding appropriate asymptotic (UV) AdS configurations.  

\begin{figure}[h]
\centerline{\epsfxsize=2.5in
\epsffile{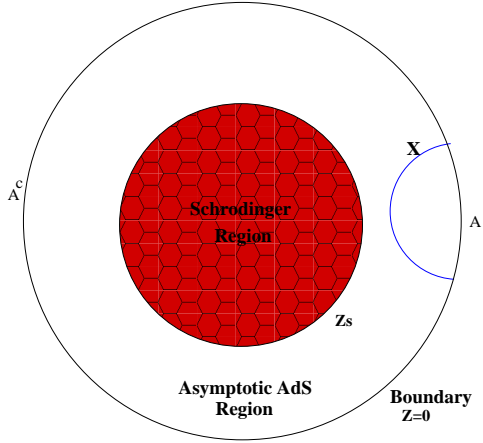} }
\caption{\label{figure4} \it The Schr\"odinger  screen is located at
$z=z_{s}$. 
The central  Schr\"odinger spacetime will not 
be accessible to an observer located in the asymptotic region ($z<z_{s}$). 
The central region would appear as a `halo' to an outside observer.}
\end{figure}

\subsection{ Schr\"odinger (spacetime) screen}
We shall selectively 
modify the Schr\"odinger
solutions \eqn{schsol3} and   attach
asymptotic AdS configuration to these IR solutions. Thus  
we  apply a  `Galilian boost' to the  Schr\"odinger solutions
along the spacial $x^-$ coordinate,
\be
x^-\to x^- - 
v x^+, ~~~~x^+\to x^+ \ .\ee
Here $v>0$ is a   boost parameter and is dimensionless. Taking $
v<0$ does not lead to any interesting situation. 
In the  boosted coordinates the metric \eqn{schsol3} 
assumes the following form
\bea\label{schsol31}
ds^2_{Sch-AdS}&=& R_p^2 z^{p-3\over p-5} \bigg[ 
\{({v\over z^2} -{\beta^2\over z^{4/p-5}}) (dx^{+})^2 +{-dx^{+}dx^{-}
+d\vec x_{(p-1)}^2\over z^2}  
+{4\over (5-p)^2} {dz^2\over  z^2} \}  + d\Omega_{(8-p)}^2 \bigg] \br
&\equiv& R_p^2 z^{p-3\over p-5} \bigg[ 
\{ K_s { (dx^{+})^2\over z^2} +{-dx^{+}dx^{-}
+d\vec x_{(p-1)}^2\over z^2}  
+{4\over (5-p)^2} {dz^2\over  z^2} \}  + d\Omega_{(8-p)}^2 \bigg] ,
\eea 
(Alternatively, one could directly
obtain them following the boosted bubble solutions given in the Appendix.)
While the dilaton and $(p+2)$-form flux will remain unchanged. 
The new function $K_s$ is given by
\be\label{schint032}
K_s= 
v -({z\over z_{IR}})^{2p-14\over p-5}\equiv
v \left(1 -({z\over z_{s}})^{2p-14\over p-5}\right)
\ee 
The various parameters are related as 
$\beta^2=z_{IR}^{14-2p\over p-5}$ and 
$v=(z_s/z_{IR})^{2p-14\over p-5}$. Note
that $K_s$ vanishes for the  special value
$z=z_{s}$ and will flip its sign across 
this point. Due to this the two  light-cone coordinates $x^+, x^-$
will  exchange their respective space and time nature across $z=z_s$. 
Thus, in the asymptotic  region ($0<z<z_{s}$),  
where $x^+$ will behave as spacelike direction with
$x^-$ being the time, while in the interior  Schr\"odinger
region ($z>z_{s}$) their roles would get reversed. 
This is somewhat reminiscent of the situation we encounter
 at the horizon of a black hole. But, the $z=z_s$ surface 
is a cool frontier, as
there is no real singularity hidden inside  the Schr\"odinger region. 
The curvature scalar remains smooth every where. 
In order to check this, we  take the case of  D3-brane, 
for which dilaton is  constant and  the string metric becomes
$$R^2 
\{({v\over z^2} -{\beta^2 z^2}) (dx^{+})^2 +{-dx^{+} dx^{-}
+d\vec x_{(2)}^2\over z^2}  
+ {dz^2\over  z^2} \}  + R^2 d\Omega_{(5)}^2  $$ 
which is a constant curvature spacetime. Note that the determinant of the string
 metric is independent of $K_s$.
 Thus in these new  coordinates 
there is  a kind of {\bf screen}
located at $z=z_{s}$, 
 which an outside (AdS) observer will always encounter  whenever it crosses 
the screen. Here onwards
we shall refer  the $z=z_s$ hypersurface, which is  topologically $R^{1,p}
\times S^{8-p}$,  simply as a 
  {\bf Schr\"odinger spacetime screen} in order to 
distinguish it from a  black hole horizon.
The interior  of  Schr\"odinger screen will always include a Schr\"odinger spacetime.
The interior spacetime would appear as a `halo' 
from  the outside of the screen; see the sketch in figure 
\eqn{figure4}.\footnote{Note that there is no obstruction for an observer 
in crossing the screen, only thing it has to do after crossing the 
screen is that it has to readjust its coordinates $(x^+,x^-)$. The $z$ remains
 a valid holographic coordinate across $z=z_s$. 
The spacetime is geodesically complete in this sense. But an static observer
located in the $z<z_s$ cannot communicate with its counterpart
 situated in $z>z_s$ region. } 
Ultimately, we think that this situation 
leads to a {\it  reduction in the entanglement entropy} as we  find next. 
The main reason for this reduction is  that the 
interior side of the screen  remains 
inaccessible  to the outside observers. Correspondingly 
in the boundary CFT some states would be `disentangled'.\footnote{ This might
appear analogous to the screening effects in condensed matter systems,
perhaps due to some concentration of hidden negatively charged (hole) states.} 

\subsection{ Lightcone compactification and IR singularity}
We discussed above that Schr\"odinger screen is a 
smooth surface where the two lightcone
coordinates exchange their spacetime nature. 
Especially in the asymptotic AdS
region,  $0<z<z_{s}$, where $x^+$ behaves as a spatial coordinate,
we could  think of compactifying  $x^+$ 
 on a circle; $x^+\sim x^+ + 2\pi r^+$.
The compactified solution, which is  good in the  asymptotic  region
only, will be given by   (in  Einstein frame metric) 
\bea\label{schint035}
&&ds^2_{p+1}
=L^2 z^{2(p^2-7p+14)\over (p-1) (p-5)}K_s^{1\over p-1} \bigg[ 
 -{(dx^{-})^2\over 4 z^2 K_s}+ {d\vec x_{(p-1)}^2\over z^2}  
+{4\over (5-p)^2} {dz^2\over  z^2}   \bigg]   
\eea 
where $x^-$ is to be treated as being the time coordinate.
There is a running  dilaton field and the Kaluza-Klein gauge field
\bea\label{conf1}
&& e^{2\phi_{p+1}}\sim {z^{5-p\over2}\over \sqrt{K_s}} \br
&& A_{(1)} \sim -{1\over 2 K_s}dx^-\eea
Note $K_s= 
v -({z\over z_{IR}})^{2p-14\over p-5}$, 
is positive definite outside the Schr\"odinger
region. Thus
the  $(p+1)$-dimensional metric 
\eqn{schint035}  
remains faithful only outside of the Schr\"odinger screen. 

It is important to note that in  the lightcone  
compactified solutions \eqn{schint035}, we have 
unwittingly introduced an essential singularity at $z=z_s$, 
because this lightcone 
circle is shrinkable at $z=z_s$. Thus the 
compactified description holds good in the $z<z_s$ region only 
and it breaks down near $z=z_s$.\footnote{ 
We would like to note that it is the usual situation whenever
compactification involves a shrinkable circle. Hence
we should either remove  or resolve this singularity. 
The simple way of handling the substringy circle size in our
 solutions would be
 to lift the solution back to ten dimensions. In 10D, 
we may  include higher order corrections to the string solutions near $z_s$,
which could cloak the singularity. But this study is beyond the scope of this
paper. However, when viewed from 10D  the compact lightcone
coordinate starts becoming null as $z\to z_s$. Eventhough this 10D
 geometry cannot be trusted near $z_s$, but DLCQ
description might eventually hold good there; see \cite{malda} 
for DLCQ of null lightcone compactification. 
Alternatively, one can  do a T-duality along the compact circle as well. 
} 
Back in ten dimensions we already
know that the `noncompact' solutions are smooth at $z=z_s$.  
The singularity gets 
introduced in an artificial way only,  when we employ
the compactification. 
In the light of this discussion we conclude that
whatever conclusions we draw in the rest of the paper
will only be good for an effective description
valid over a limited  UV regime $z<z_s$.

\subsection{ Confinement in the IR region}

As we learned above, upon lightcone
 compactification the $(p+1)$-dimensional effective string
coupling in \eqn{conf1}, $<e^{\phi_{p+1}}>$,  
 becomes stronger near $z=z_s$ much faster than
the usual (relativistic) RG flow. In fact it becomes 
singular, so we cannot trust the 
lower dimensional interpretation very close to $z=z_s$. 
It would  be appropriate that as we approach $z=z_s$
we should ideally up-lift the solution \eqn{schint035} 
back to 10-dimensions and consider stringy coorections.

The presence of the KK fields in \eqn{conf1} implies 
a presence of finite chemical potential, 
and following from eq.\eqn{kju1} we can determine 
$$\mu \sim  -{1\over 2 r^+ v}$$ 
in the $p$-dimensional
Schr\"odinger CFT near the UV
fixed point, associated with an effective (negative) 
charge density 
$$\rho \sim -{(R_p)^{3p-6}(r^+)^2 \over  G_{p+2}}
\left({1\over z_{IR}}\right)^{2p-14\over p-5}.$$
Only  change is in the sign of the charge density from the 
expressions given in \eqn{kju1} for the Lifshitz backgrounds.
Thus these quantities make the defining 
characteristics of the boundary Schr\"odinger CFT near 
the UV fixed point. 
It means  that such a Galilean
Yang-Mills theory would flow towards the IR region where its 
effective gauge coupling becomes stronger
near an intermediate scale $z= z_s$ and the bulk  geometry would 
ultimately encounter a Schr\"odinger screen. 
Some of these are the known characteristics of 
confining YM theories, more like confinement in  QCD. 
(For  example, we could consider $N$ D4-branes with Schr\"odinger 
 IR deformation so that the corresponding boundary theory 
is  4D Schr\"odinger CFT, i.e. a lightcone compactified
5D SYM with suitable operator deformations 
related to chemical potential and  (negative)
charge density as described above.)  
The  generic analysis  based on  Schr\"odinger-AdS bulk spacetime
predicts a confinement in  
$D\le 4$ for a Schr\"odinger CFT at some intermediate IR scale $z_s$.

\subsection{Entanglement Entropy in Schr\"odinger-AdS systems}

In order to find the entanglement entropy (EE) of a subsystem 
of a Schr\"odinger CFT, we will use the 
interpolating solution  
 \eqn{schint035}, only  for simplicity.\footnote{ Note, the entanglement
entropy can 
 also be obtained by using the 10D metric
where lightcone coordinate is not compactified. 
The basic expression remains unchanged for a strip-like subsystem except minor
changes in the Newton's constant and volume like parameters.} Thus
we  pick up a subsystem $A$ (with its boundary $\partial A$) 
in the  CFT. The subsystem $A$ will  naturally
have an entanglement of its states
with its complement  $A^c$, which  supposedly is  comparatively  larger system.
The  entanglement entropy of the subsystem $A$ 
 is described  geometrically (holographically) 
in terms of the area of an extremal surface $X_{(p-1)}$ 
(spacelike $(p-1)$-dimensional surface) ending on  
the boundary $\partial A$, see \cite{RT}. Thus we have for entanglement entropy
\be S_{Ent}(A)={1\over 4 G_{p+1} }[Area]_X
\ee 
The extremal surface $X_{(p-1)}$ lies  within the bulk spacetime,
which is a $(p+1)$-dimensional 
Schr\"odinger spacetime being asymptotically conformally  AdS$_{p+1}$. 
We pick up the subsystem $A$ such that it is  a rectangular strip
with width along $x^1(z)$ and stretched along rest of $ x^i$'s, 
at any fixed time. 
(Note  we have taken $x^+$ being  a compact coordinate.) 
Also, $x^-$ is being identified with boundary
time coordinate. The range of $x_1$ coordinate is 
$-l/2 \le x_1\le l/2$ and the regulated (but large) size of
other coordinates is $0\le x^i\le l^i ~(l^i\gg l)$ for $i=2,...,p-1$.
Considering the $(p+1)$-dimensional Einstein metric \eqn{schint035},
 we find 
\be\label{schkl1}
S_{Ent}={V_{p-2}L^d\over 2 G_{p+1}}
\int_{z_\ast}^{z_\infty}dz ~ z^{9-p\over p-5} 
\sqrt{K_s}\sqrt{{4\over (5-p)^2} + (x_1')^2}
\ee  
where $z_\infty\approx 0$ is the UV cut-off and $z_\ast$ is  the turning point.
$V_{p-2}=l_2\cdots l_{p-1}$ is the volume of the ensemble box stretched along the spatial
directions $x_2,\cdots, x_{p-1}$.   
$K_s$ is as given in \eqn{schint032}. From \eqn{schkl1} we determine that  
the minimal surface satisfies the first order equation
\bea\label{kl3}
{dx_1\over dz}&=& {2\over 5-p} {C z^{p-9\over p-5}\over
\sqrt{v
-({z\over z_{IR}})^{2(p-7)\over p-5}
-C^2 z^{2(p-9)\over p-5}}} \br
&\equiv & {2\over 5-p} { ({z\over z_c})^{p-9\over p-5}\over
\sqrt{
1
-({z\over z_s})^{2(p-7)\over p-5}
-({z\over z_c})^{2(p-9)\over p-5}}} \br
\eea 
where $C$ is an integration constant and we redefined ${C^2\over v}=
({1\over z_c})^{2(p-9)\over p-5}$. At the turning point
\be
1
-({z\over z_s})^{2(p-7)\over p-5}
-({z\over z_c})^{2(p-9)\over p-5}=0
\ee
where 
 $x_1'|_{z_\ast}=\infty$ and $x_1(z_\ast)=0$. 
While at the boundary points 
$x_1'|_{z=0}\sim 0$, where boundary condition is $x_1(0)=l/2$. 

Finding solutions of the first order
differential equation \eqn{kl3} is much like solving a classical orbit in
 the central force problem with the given initial conditions. 
The term  $C^2 z^{2(p-9)\over p-5}$ plays the role of a repulsive 
`centrifugal type potential', while the term
$({z\over z_{IR}})^{2(p-7)\over p-5}$ behaves like a {\it repulsive} 
central potential. It is easy to see that the latter repulsive
force arises due to the interior Schr\"odinger region which  repels 
the classical trajectories approaching from the boundary 
 with certain fixed  energy 
 $E\equiv E(v)$. Higher is the $v$ the higher will be the penetrating power of the 
projectile, which ultimately has to bounce back. 
In any case these orbits can never penetrate the central potential barrier. 
The point of closest approach, $z_\ast$, will be restricted to being $z_\ast > z_s$.
Thus the IR Schr\"odinger spacetime   leads to
 a central repulsive force, while the universal component of 
repulsive force 
comes from the curvature of AdS spacetime. Note that this situation is 
 different to that of the Lifshitz solutions
discussed in the last section,
 where there existed an `attractive' component in the effective potential, see \cite{hs13}. 
 
\begin{figure}[h]
\begin{minipage}[t]{3in}
\centerline{\epsfxsize=2.5in
\epsffile{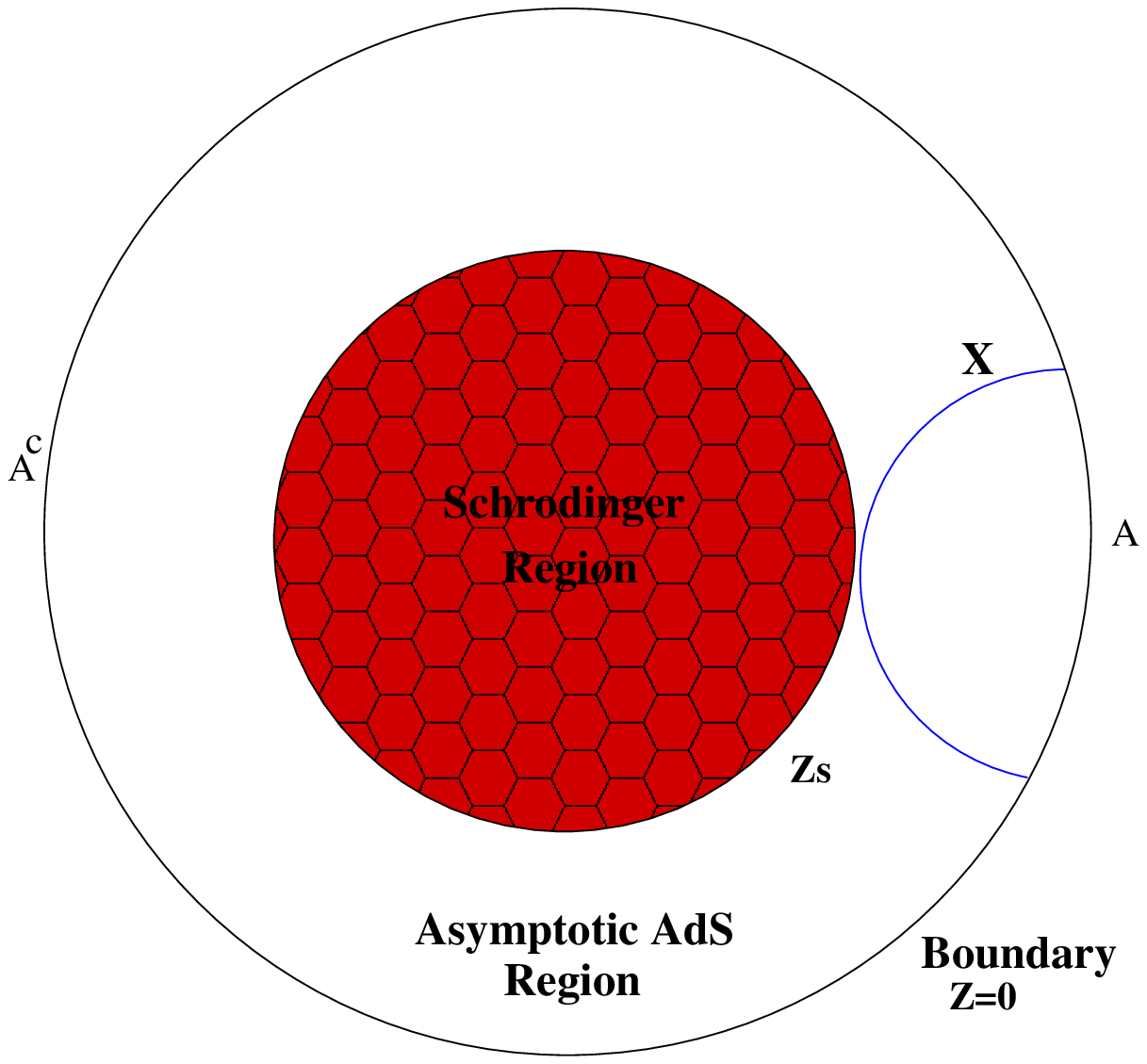} }
\caption{\label{figure5}\it
A large  extremal surface $X$ will glance the 
Schr\"odinger  screen but  at the safe distance
$z_{\ast}< z_{s}$. }
\end{minipage}
\hspace{.4cm}
\begin{minipage}[t]{3in}
\centerline{\epsfxsize=2.5in
\epsffile{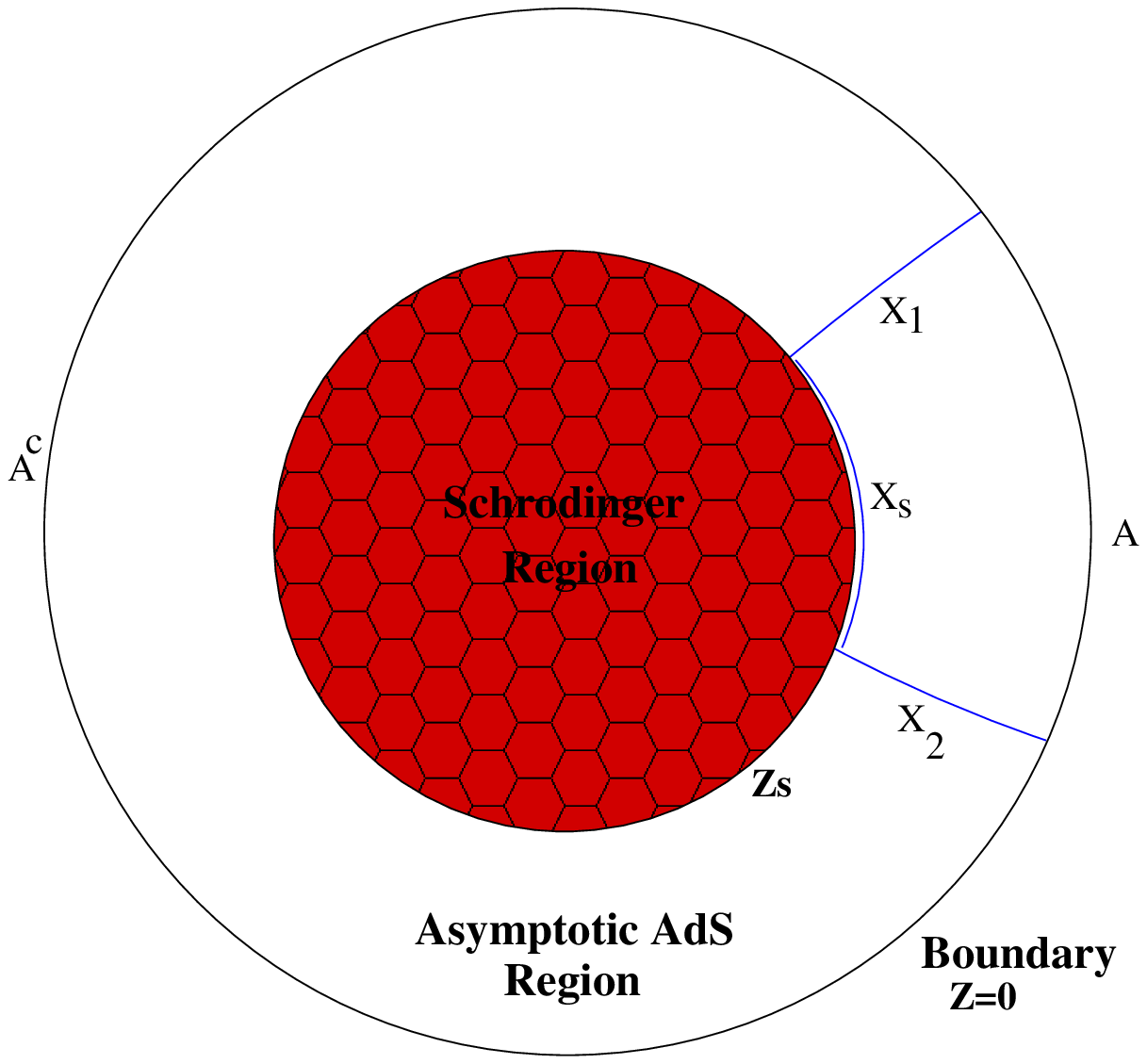} }
\caption{\label{figure6} \it
Eventually as the size of the boundary subsystem  increases, the 
extremal surface has three parts $X_1,X_s$, and $X_2$. The $X_s$ part glances the 
Schr\"odinger  screen from a safe location
$z=z_{\ast}$ . }
\end{minipage}
\end{figure}

We shall be interested in the situation where $z_s \gg z_c$.
It can be inferred that in the Schr\"odinger case the turning 
point will always arise in the region $z<z_c$ and in this region 
 the geometry \eqn{schint035} is well defined. 
While in the Lifshitz case, 
due to attractive component in the potential the turning point  
 arises for $z >z_c$. So keeping everything else the same,
the Lifshitz orbits are generally longer in depth.   
This gives  an entanglement
 entropy formula for the Schr\"odinger system as
\be\label{schfg3}
S_{Ent}={V_{p-2}L^d\over 2 G_{p+1}}
{2\sqrt{v}\over (5-p)} 
\int_{z_\ast}^{z_\infty}dz ~z^{9-p\over p-5} 
{1-({z\over z_{s}})^{2(p-7)\over p-5}
\over\sqrt{ 1
-({z\over z_s})^{2(p-7)\over p-5}
-({z\over z_c})^{2(p-9)\over p-5}
  }}
\ee  
If we set $z_{s}=\infty$, the expression 
\eqn{schfg3} 
reduces to the entanglement entropy in the relativistic CFT system.
The turning point of the extremal surface
 in the conformally AdS case appears at the value
$z=z_c$. 
While we will always get $z_\ast < z_c$ 
for the Schr\"odinger-AdS system due to repulsive nature of the potential. 
Thus the area of the extremal surface $X$ 
is going to be smaller for the Schr\"odinger-AdS spacetime compared to 
the purely conformally AdS case.
While  we already know that the area of the extremal surface has been larger
in the Lifshitz-AdS case as compared to the pure conformally AdS case \cite{hs13}. 
Thus we can establish the hierarchy of the entanglement entropies,
\be
S_{Ent}^{Lif-AdS} > 
S_{Ent}^{AdS}>S_{Ent}^{Sch-AdS}  .
\ee
provided the global system parameters, like $v$, the size $l$ (or $z_\ast$) and  $z_{IR}$ 
 are kept the same.

\begin{figure}[h]
\begin{minipage}[t]{3in}

\centerline{\epsfxsize=2.5in
\epsffile{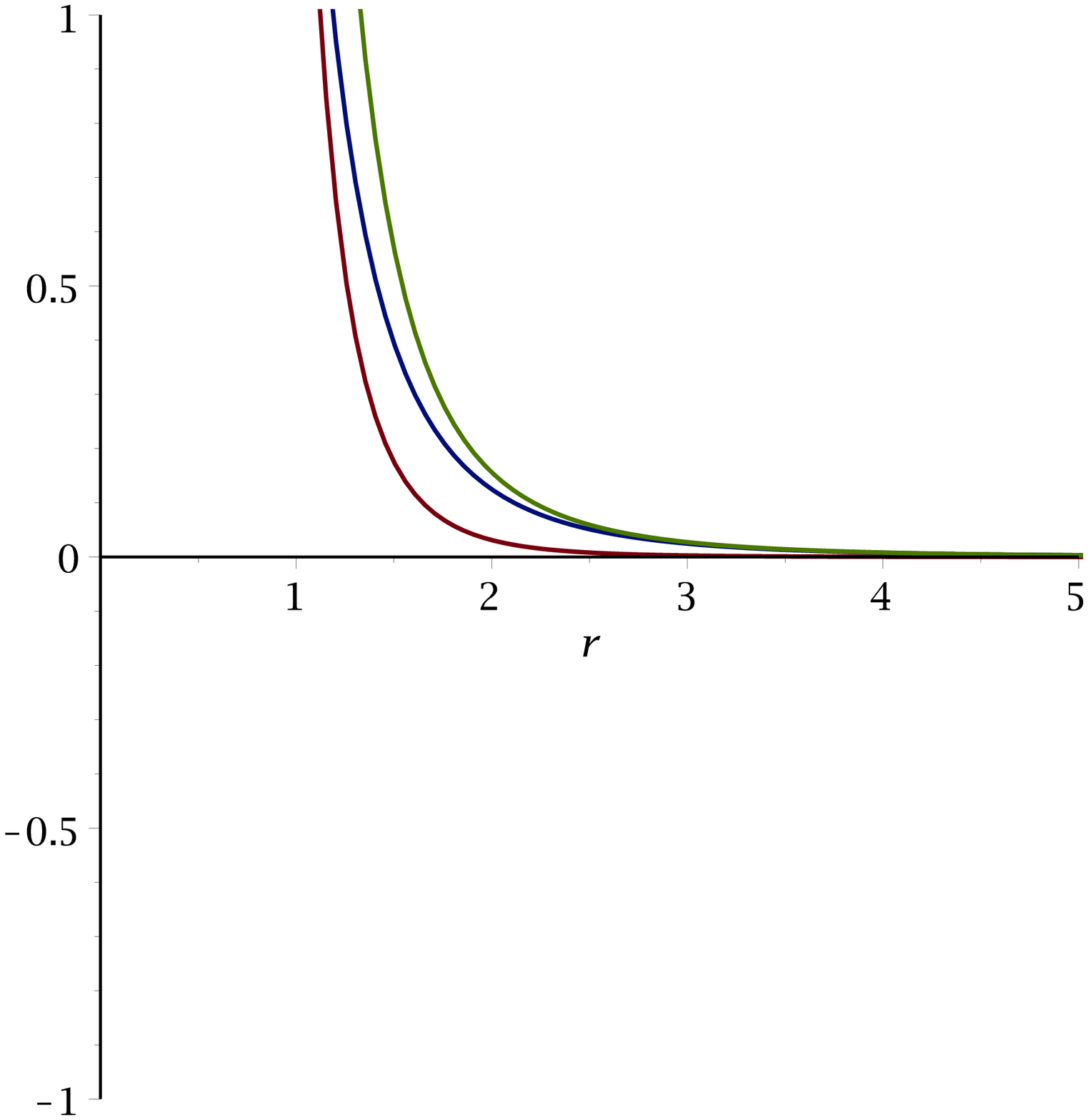} }
\caption{\label{figure7} 
\it A  plot of  various components of the
effective (Schr\"odinger-AdS) potential $V_{eff}$ (for $p=3$). 
The green color (right most) 
curve gives the resulting potential which  is repulsive.  }
\end{minipage}
\hspace{.4cm}
\begin{minipage}[t]{3in}
\centerline{\epsfxsize=2.5in
\epsffile{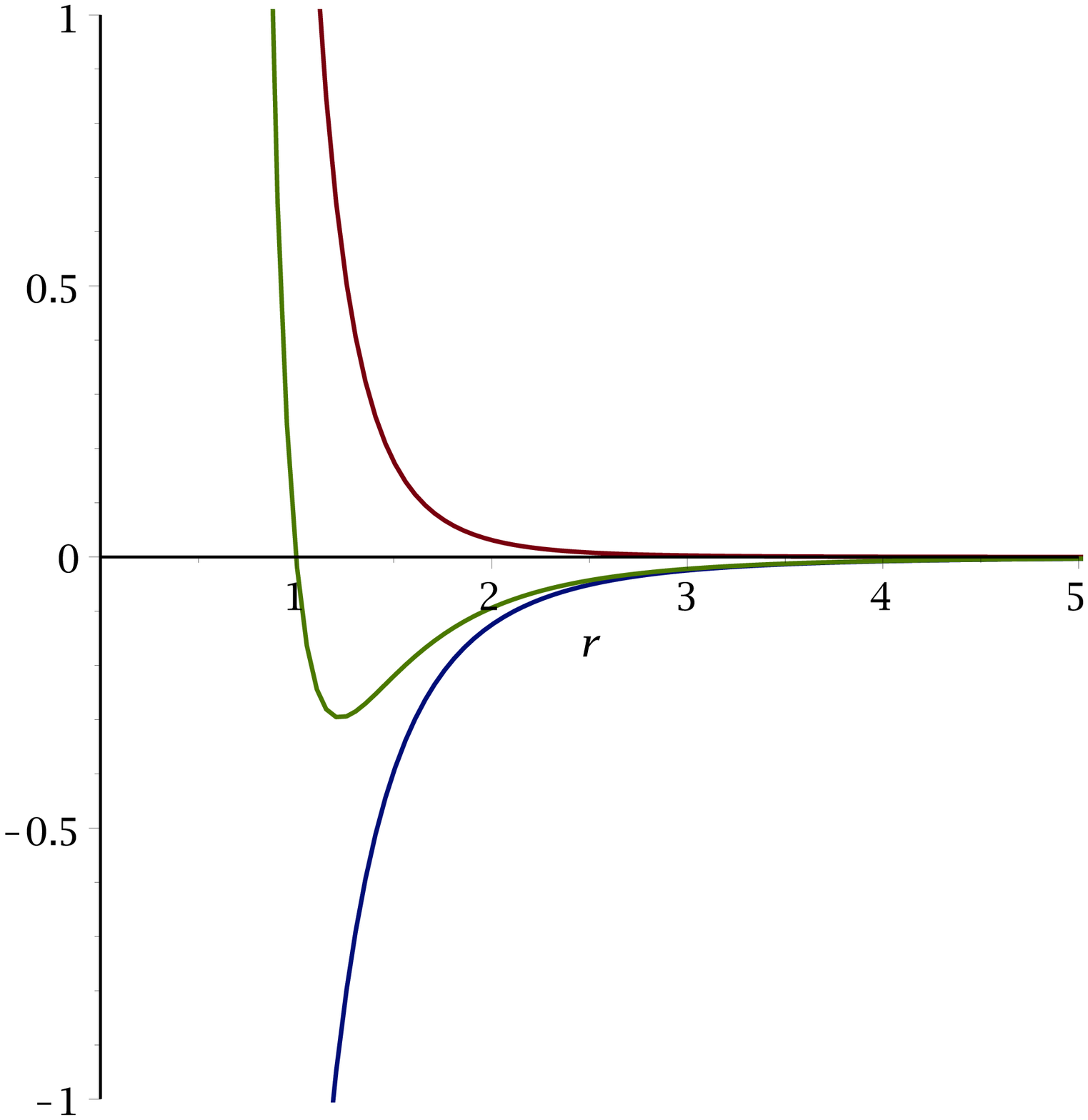} }
\caption{\label{figure8} 
\it A  plot of the various components of the 
 Lifshitz-AdS potential $V_{eff}$, for $p=3$. 
The green color (middle curve) 
 gives the resultant potential which is attractive in the
asymptotic regime.  }
\end{minipage}
\end{figure}

\subsection{A troika of the solutions} 
The analysis of the previous sections provides us with  a general entanglement
 entropy formula for the Schr\"odinger-AdS , conformally AdS and 
 Lifshitz-AdS bulk solutions
\be\label{schfg3d}
S_{Ent}={V_{p-2}L^d\over 2 G_{p+1}}
{2\over (5-p)} 
\int_{z_\ast}^{z_\infty}dz ~z^{9-p\over p-5} 
{ K(z)
\over\sqrt{ v-V_{eff}}}.
\ee  
The  effective potential
appearing in the three cases can be classified as
\bea
V_{eff} &=&  -({z\over z_{IR}})^{2(p-7)\over p-5}
+C^2 z^{2(p-9)\over p-5}~~~~~{\rm Lifshitz-AdS} \cr
&=&({z\over z_{IR}})^{2(p-7)\over p-5}
+C^2 z^{2(p-9)\over p-5}~~~~~{\rm Schrodinger-AdS} \cr
&=&
C^2 z^{2(p-9)\over p-5}~~~~~~~~~~~~~~~~~~~~~~~{\rm conformally ~AdS} 
\eea
while  the numerator in the integrand will read as 
\bea
K(z) &=& v +({z\over z_{IR}})^{2(p-7)\over p-5}
~~~~~{\rm Lifshitz-AdS} \cr
&=&v-({z\over z_{IR}})^{2(p-7)\over p-5}
~~~~~{\rm Schrodinger-AdS} \cr
&=&v ~~~~~~~~~~~~~~~~~~~~~{\rm conformally ~AdS} 
\eea
Especially for
$p=3$ case  the effective potential is  plotted for some arbitrary but fixed
values of the parameters
in the figure \eqn{figure7} for  Schr\"odinger-AdS,
 and in figure \eqn{figure8} for the Lifshitz-AdS  solutions. 
Note that  we have also redefined the holographic
coordinate back to $r=z^{2\over p-5}$ for the convenience in these plots, so that 
\bea
V_{eff} 
&=&  -({r_{IR}\over r})^{7-p}
+{C^2\over r^{9-p}} ~~~~~{\rm Lifshitz-AdS} \cr
&=&  ({r_{IR}\over r})^{7-p}
+{C^2\over r^{9-p}}
~~~~~{\rm Schrodinger-AdS} \cr
&=&  
{C^2\over r^{9-p}}
~~~~~~~~~~~~~~~~~~{\rm conformally~ AdS} 
\eea

\section{Numerical analysis}
Although we have qualitatively understood the holographic picture of 
the  Schr\"odinger-AdS  bulk solutions,
it would be worth while to make it concrete with some numerical
calculations. For this purpose  we pick up an special case of D3-brane. 
We rewrite 
the integral equation governing the extremal (entanglement) surface as
\bea\label{num1}
x_1(b)= r_\ast^3 \sqrt{K_\ast}\int^{b}_{r_\ast}dr {r^{-5}\over
\sqrt{1- {r_s^4\over r^4}- {r_\ast^6 K_\ast\over r^6}}}
\eea  
note that $x_1(r_\ast)=0$. We have taken $K(r)=1-{r_s^4\over r^4}$ and in our
 notation $K_\ast\equiv K(r)|_{r=r_\ast}$. Actually the parameter 
$b$ representing the boundary value
of holographic  coordinate
should be taken  reasonably large so as to represent boundary location. 
Similarly the entanglement 
entropy integral can  be expressed as
\bea\label{num2}
S(b)= S_0 \sqrt{v} \int^{b}_{r_\ast}dr {r^{-1}K(r)\over
\sqrt{1- {r_s^4\over r^4}- {r_\ast^6 K_\ast\over r^6}}}
\eea
where $S_0\equiv {V_{1}L^2\over 2 G_{4}}$ is an overall 
constant in the entropy expression. We will set $S_0=1$ in 
rest of the analysis.
In our strategy we  consider a  small (perturbative)
value of $r_s=.1$ and  set  $v=100$. Note that turning point of the extremal surface
for Schr\"odinger-AdS solutions
always arise outside of the screen, that is $r_\ast > r_s$. For 
various values of these turning points, $r_\ast>r_s$,  we numerically 
evaluate above first integrals. 
The  graphs for $x_1(b)$  vs $b$ 
have been  plotted in figure \eqn{figure12} over sufficiently large range of $b$. We have 
selectively  taken $r_\ast=.12,~ .15,~ .2$ and $.5$. 
In fact one could  
take any other set of allowed values. We observe that $x_1(b)$ 
graphs become flatter at large $b$, which gives the idea of system size. 
Recall  that 
typically the asymptotic value  $x_1(\infty)$ gives us the 
system size, $l$, or the boundary value. Also we observe as the turning point
$r_\ast$ becomes deeper and 
deeper the subsytem size in the boundary CFT increases. This is 
an expected behaviour. {\it The subsystem size will be largest for 
the trajectory
which passes by near most to the Schr\"odinger screen at $r=r_s$.}
We note that, as the size of system $A$ increases, a part of 
extremal surface, $X_s$, would  start glancing the 
Schr\"odinger  screen from a  fixed location
$r=r_{\ast}$, see the figure \eqn{figure6}. Note however
these trajectories can never touch the
screen located at $r=r_s$.
The entanglement entropy contribution
from $X_s$ part of the surface, namely the part parallel to the 
Schr\"odinger screen, would
  however come out to be vanishingly small, due to $K^{1\over p-1}$ factor in the metric. 
This implies that there is a limit to the entanglement entropy. Thus after a point,
even if the subsystem size $l$ increases, the entropy would stop increasing 
any further. Hence there is a saturation point in the Schr\"odinger-AdS system. 
We  note down some of the values of entanglement entropy (for which we have safely 
taken UV cut-off to be $b=1.5$)
\be\label{num4}
S^{Sch}_{r_\ast=.12}=11.1817,~~~
S^{Sch}_{r_\ast=.15}=11.1784,~~~
S^{Sch}_{r_\ast=.2}=11.1509,~~~
S^{Sch}_{r_\ast=.5}=10.705,~~~
\ee
Thus we have 
\be\label{num5}
S^{Sch}_{r_\ast=.12}>
S^{Sch}_{r_\ast=.15}>
S^{Sch}_{r_\ast=.2}>
S^{Sch}_{r_\ast=.5}\ee
 which means that the entanglement entropy increases alongwith 
the subsystem size. But as the extremal surfaces get closer to the
Schr\"odinger screen, the entropy of the systems stops increasing 
with the system size. 
For example, for  the extremal surfaces with
$r_\ast=.12$ and $r_\ast=.15$  there is no  appreciable 
change in the entanglement entropy. {\it We conclude that the part 
of the extremal surface, $X_s$, parallel to the 
Schr\"odinger screen has negligible contribution to the entanglement 
entropy}. That is, {\it  correspondingly some states in the CFT  would have 
negligible contribution to the entanglement} or will remain hidden. 
Thus the Schr\"odinger screen leads to {\it reduced entanglement} 
in  Schr\"odinger like solutions
embedded in an asymptotically (conformally) AdS spacetimes. 

\begin{figure}[h]
\begin{minipage}[t]{3in}
\centerline{\epsfxsize=2.5in\epsffile{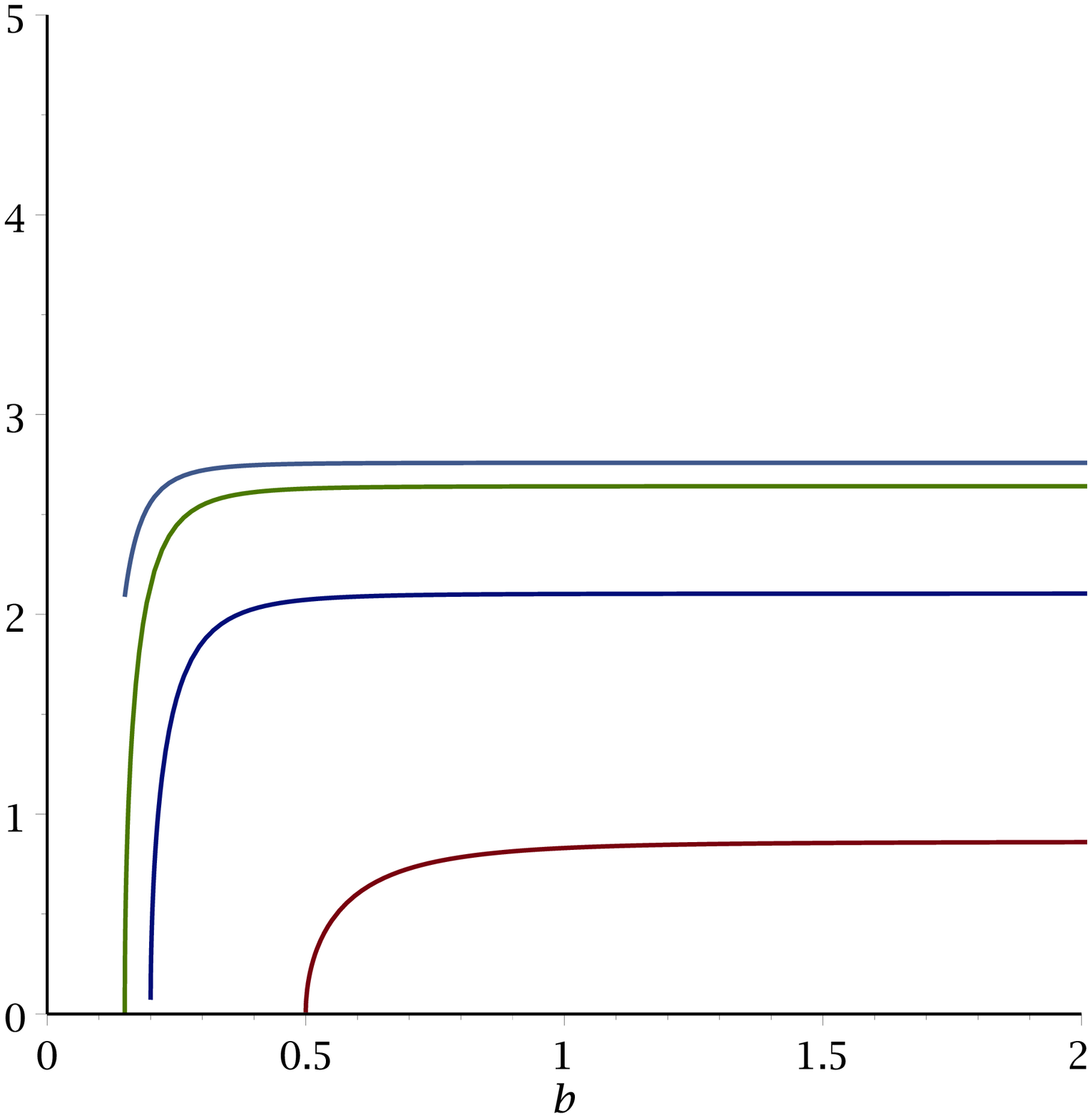} }
\caption{\label{figure12} 
\it Plots of the extremal trajectories, $x_1(b)$ for the Schr\"odinger CFT
for the turning points $r_\ast=.12, .15, .2$ and $.5$. It shows
that closer is the turning point to $r_s=.1$ (the Schr\"odinger screen) the
larger is its impact parameter (or the size of the boundary CFT subsystem).  }
\end{minipage}
\hspace{.4cm}
\begin{minipage}[t]{3in}
\centerline{\epsfxsize=2.5in\epsffile{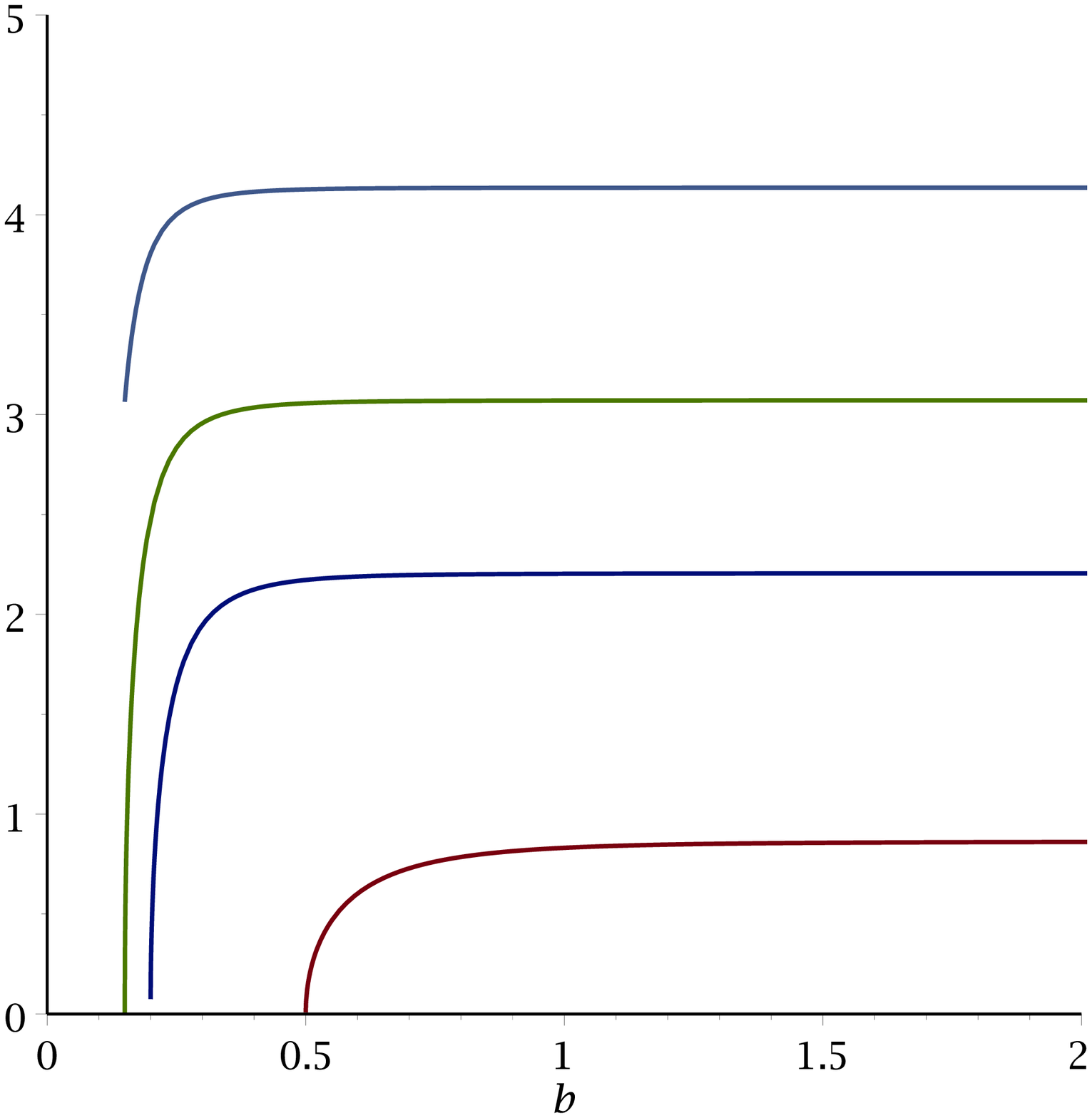} }
\caption{\label{figure15} \it 
Plots of the extremal trajectories $x_1(b)$ for the Lifshitz CFT
for the turning point values $r_\ast=.12, .15, .2$ and $.5$. It shows
that smaller the value of $r_\ast$
larger is the impact parameter (or size of the boundary subsystem).  }
\end{minipage}
\end{figure}

This is altogether very different situation when compared  with 
the Lifshitz-AdS cases.
The first  major difference in the two solutions is that the Lifshitz 
solutions do not have a screen (but they can include a black hole horizon). 
So, in principle, a large extremal surface can have a turning point situated 
well inside  the  Lifshitz (IR) region. 
For the purpose of the comparison, 
we have taken the same
values of the turning points as  in the Schr\"odinger case above. 
The  graphs for $x_1(b)$  vs $b$ 
have been  plotted in figure \eqn{figure15} for the Lifshitz-AdS case. We have 
selectively  taken $r_\ast=.12,~ .15,~ .2$ and $.5$. As expected orbits are large
in the Lifshitz-AdS case.
We do also find the entanglement entropy
(with the same value $b=1.5$ and other parameters being the same)
\be\label{num6}
S^{Lif}_{r_\ast=.12}=11.253, ~~~
S^{Lif}_{r_\ast=.15}=11.224, ~~~
S^{Lif}_{r_\ast=.2}=11.176, ~~~
S^{Lif}_{r_\ast=.5}=10.709, ~~~
\ee
From the eqs. \eqn{num4} and \eqn{num6} we  conclude that in general 
\be
 S^{Lif-AdS}> S^{AdS} >S^{Sch-AdS} .
\ee

\section{Conclusion}
We have studied  the Schr\"odinger type nonrelativistic 
D$p$-brane solutions having  conformally
AdS asymptotic geometries. 
These specially engineered Schr\"odinger-AdS solutions  possess a smooth
`horizon' like hypersurface,  more appropriately a screen (or a mask),
across which lightcone coordinates $(x^-,x^+)$ flip their respective
space and time nature. We only discussed zero temperature solutions. 
As we know a black hole horizon 
 behaves like a finite temperature 
frontier and whose area gives  black hole entropy.
So in this sense the {\em Schr\"odinger screen} is totally different from
the Schwarzschild  horizon. 
For the Schr\"odinger-AdS solutions,
 an observer situated in the asymptotic AdS region will find the interior 
Schr\"odinger region to be completely  inaccessible. {\it
Thus some information will get hidden behind the Schr\"odinger screen}. 
 
The presence of asymptotically AdS region in the
Schr\"odinger solutions  
 requires us to include  finite chemical potential
and a (negative) charge density.
We have  estimated the entanglement entropy of the strip-like subsystem in 
the boundary. It is found that the masked 
Schr\"odinger region does not contribute to the entanglement and this
results in a net reduction of the entanglement entropy.  The net  effect of 
the Schr\"odinger screen is that it divides the spacetime into two isolated parts
and this results in a loss of quantum information. This may indicate that 
a finite concentration of the negatively charged fermions,  perhaps a
finite concentration of the hole states (or vacancies in the fermi surface) 
could reduce the entanglement. We have worked in the regime where
the charge density is much smaller (a perturbative regime) 
than the finite effect of chemical potential in the CFT. 
We do find that as the   subsystem size increases the entanglement entropy
 initially increases along with the system size but it
does not increase beyond a point and  gets saturated. 
On the physical grounds this might be due to the fact that 
screening effects of the Schr\"odinger
screen come into play.  It requires to further investigate these kind of
solutions and also identify analogous
quantum phenomenon in condensed matter physics or fluid systems.

\vskip.5cm
\noindent{\it Acknowledgements:} The author is an  Associate of the ICTP, Trieste
and acknowledges the generous  support from the centre, where 
this work has come to the completion.  

\vskip.5cm
   
\appendix{
\section{The D$p$-branes and the hyperscaling}
The maximally supersymmetric near horizon  AdS  solutions  are 
given by \cite{itzhaki}
\bea\label{sol2de1}
&&ds^2_{AdS}=R_p^2r^{p-3\over2}\bigg[ 
r^{5-p}\{  -dx^{+}dx^{-}
+d\vec x_{(p-1)}^2\}  
+{dr^2\over  r^2}  + d\Omega_{(8-p)}^2 \bigg]\ , \br
&& e^\phi=(2\pi)^{2-p}g_{YM}^2 
R_p^{3-p} r^{(7-p)(p-3)\over4} 
\eea 
along with a suitable $(p+2)$-form field strength
\be
F_{p+2}=(7-p)R_p^{2p-2}r^{6-p} 
dr\wedge dx^+\wedge dx^-\wedge [dx_{(p-1)}]\ee
for the electric type D$p$-branes $(p<3)$ and a $(8-p)$-form
\be F_{8-p}=(7-p) R_p^4\, \omega_{8-p}\ee
for the magnetic type (Hodge dual) D$p$-branes $(p>3)$.
Specially for D3-brane case
we  have $F_5=4 (1+\star)\omega_5$, which is self-dual 5-form field strength. 
We have introduced $x^+,x^-$ as  lightcone coordinates along the 
world volume of the branes,  
and $\vec x_{(p-1)}$ represents other $(p-1)$ spatial directions parallel 
to the D$p$-brane, and as usual $r$ is the radial (holographic) coordinate.
The interpretation of various parameters can be found in 
\cite{itzhaki} and also given in  \cite{hs10a}.

As emphasized in \cite{hs13}, in the above  conformally  $AdS_{p+2}\times S^{8-p}$ 
solutions, we could have  taken  slightly modified  $AdS$ line element 
\bea\label{sol2de2}
&&ds^2_{AdS}=R_p^2r^{p-3\over2}\bigg[ 
r^{5-p}[ v (dx^{-})^2 -dx^{+}dx^{-}
+d\vec x_{(p-1)}^2]  
+{dr^2\over  r^2}  + d\Omega_{(8-p)}^2 \bigg]\ , \br
\eea 
Namely we have introduced a constant  
$g_{--}>0$ component, but note  that $v>0$. Doing this is actually harmless 
as it still remains an AdS vacua. (The reason is that the constant $g_{--}$ 
term can be reabsorved by a coordinate shift, like
$x^+\to x^+ + v x^-$, if and when  the need arises.) However, certain 
global symmetries of the metric,
such as the  lightcone boost $x^-\to \lambda x^-,~
x^+\to {1\over \lambda} x^+$, may not be manifest in the 
shifted coordinate frame. The inclusion of constant $g_{--}$ 
component in these solutions is
 useful in the following way. 
We shall be considering  (nonrelativistic)
Schroedinger-like solutions which  have nontrivial $g_{++}$ deformations.
Since Schr\"odinger solutions have $g_{++} <0$, so
  $x^+$ is a timelike coordinate
in them. Once  
 we employ the above constant shift in these Schrodinger solutions
then in some spacetime region 
 we  will find   $x^+$ to behave like a space like coordinate, 
i.e. $g_{++} \ge 0$. In this region
 we can also make it a compact direction and the subsequent 
DLCQ description of the holographic theory would then follow. 
Not that we need  large discrete momentum
 modes in the compact direction as it is 
this sector which tends to behave nonrelativistically. 

Let us  redefine the radial coordinate 
\be 
r^{p-5}=z^2 ~~~~~~~~~~{\rm for}~ p\ne 5
  \ee
With $z$ as holographic coordinate and some 
scaling of the brane coordinates the above solutions 
can be brought to the form 
\bea\label{sol2d03}
&& ds^2=R_p^2 z^{p-3\over p-5} \bigg[ 
\{ {v(dx^{+})^2\over z^2} +{-dx^{+}dx^{-}
+d\vec x_{(p-1)}^2\over z^2}  
+{4\over (5-p)^2} {dz^2\over  z^2}\}  + d\Omega_{(8-p)}^2 \bigg] \br
&& e^\phi=(2\pi)^{2-p}g_{YM}^2R_p^{3-p}{ z^{(7-p)(p-3)\over 2( p-5)}}
\eea 
along with the $(p+2)$-form flux.
One can find that under the  dilatations $z\to \xi z $,
 the  brane coordinates would rescale as
\be\label{d1}
x^{\pm}\to \xi x^{\pm}, ~~~\vec x\to \xi \vec x
\ee
while  the dilaton and 
the string metric  in \eqn{sol2d03} conformally rescale as
\bea\label{d2}
g_{MN}\to \xi^{p-3\over p-5} g_{MN},
~~~e^\phi\to \xi^{(7-p)(p-3)\over2(p-5)}e^\phi
\eea
Note this overall conformal rescaling is  
the standard  Weyl rescaling behaviour, 
 of conformally AdS solutions \cite{itzhaki}, 
giving rise to the RG flow in the boundary CFT. 
From  Eq.\eqn{d1} the  dynamical exponent of time 
is $a\equiv a_{rel}=1$, 
so that the boundary theories
 are  $(p+1)$-dimensional
`relativistic' CFT$_{(p+1)}$ with sixteen supercharges. 
Note, once $x^+$ is taken to be a coordinate on a circle, the 
boundary CFT becomes a DLCQ theory and is a $p$-dimensional theory. 
While the compactification of the bulk solution 
\eqn{sol2d03} along $x^-$ and $S^{8-p}$, 
 results in $(p+1)$-dimensional 
(Einstein) metric given as
\bea\label{sol2d}
ds^2_{p+1}
&\sim& z^{({p-5\over p-1}+{p-3\over p-5})} \bigg[ 
 -{(dx^{+})^2 \over z^2}+{d\vec x_{(p-1)}^2\over z^2}  
+{4\over (5-p)^2} {dz^2\over  z^2}   \bigg] \br 
&= &z^{2(p^2-7p+14)\over (p-1) (p-5)} \bigg[ 
 -{(dx^{+})^2 \over z^2}+{d\vec x_{(p-1)}^2\over z^2}  
+{4\over (5-p)^2} {dz^2\over  z^2}   \bigg] 
\equiv z^{2\theta \over d} 
 ds^2_{AdS_{p+1}}\ .  
\eea 
From where we can read the hyperscaling parameter to be
\be\label{thea1}
 \theta={p^2-7p+14 \over p-5}\equiv \theta_{rel} .
\ee
Note that, $d\equiv p-1$  gives the total number of spatial directions
of the boundary CFT$_{p}$. 
Let us  mention here that there is also  
a running $(p+1)$-dimensional dilaton field
\be
 e^{-2\phi_{(p+1)}}\sim z^{p-5\over2}
\ee
 as well as 
 fields arising out of the reduction of $(p+2)$-form RR field strength. 
These  solutions are  extremal solutions.

\section{Boosted Bubble solutions}
 We do note down the boosted bubble solutions for the sake of completeness.
The procedure is the same as described in \cite{hs12}. We take a AdS-bubble
solution, make a shift of the spatial lightcone coordinate, and then
Lorentz boost the system. The resultant solution is,

\bea\label{schint033s}
ds^2_{Bubble}&=&R_p^2 z^{p-3\over p-5} \bigg[
\{-{g(dx^{+})^2\over4 z^2 K}+
{d\vec x_{(p-1)}^2\over  z^2}
+{4\over (5-p)^2} {dz^2\over g z^2}\}
+{K\over z^2}(dx^- - A)^2
  + d\Omega_{(8-p)}^2 \bigg]\nonumber \\
\eea
where 1-form $$A\equiv {(1+g)-v\lambda^{-2}(1-g)\over 4K}dx^{+}$$
and the harmonic functions
 \bea\label{int034s}
g(z)&=&1-\left({z\over z_{0}}\right)^{2p-14\over p-5}\br
K(z)
&=& v-{(v\lambda^{-1}+\lambda)^2\over 4}\left({z\over z_{0}}
\right)^{2p-14\over p-5}
\equiv v-{1\over 4}\left({z\over z_{IR}}\right)^{2p-14\over p-5}.
\eea
The dilaton and the $(p+2)$-form field strength remain as in the AdS-bubble
case. The bubble solutions are the
nonextremal examples. These are smooth boosted
bubble $p$-brane solutions and the holographic coordinate range is 
given as $z_0\ge z \ge 0$. If $v\ne 0$, they become asymptotically conformally
AdS solutions.
The double limits
$1/z_0\to 0,~\lambda\to\infty$, keeping $z_{IR}={\rm fixed}$, will 
give us  Schr\"odinger-AdS solutions \eqn{schsol31}, which are extremal cases. 

}


\vskip.5cm


\begin{thebibliography}{99}

\bibitem{son} D.T. Son, 
\prd{78}{2008}{046003}, \hepth{0804.3972}.
\bibitem{bala}
K. Balasubramanian and J. McGreevy, \prl{101}{2008}{061601}, \hepth{0804.4053};
K. Balasubramanian and J. McGreevy, JHEP 01 (2001) 137, \hepth{1007.2184}. 


\bibitem{kachru} S. Kachru, X. Liu and M. Mulligan, 
"Gravity duals of Lifshitz-like Fixed Points'', 
\prd{78}{2008}{106005}, \hepth{0808.1725}.


\bibitem{malda} J. Maldacena, D. Martelli and Y. Tachikawa, 
JHEP 0810 (2008) 072, arXiv:0807.1100. 

\bibitem{nara1}
K. Balasubramanian and K. Narayan, JHEP {\bf 1008}, 014 (2010),
 \hepth{1005.3291}.
 

\bibitem{hs10}
  H.~Singh,
  ``Special limits and non-relativistic solutions,''
  JHEP {\bf 1012}, 061 (2010)
  [arXiv:1009.0651 [hep-th]].

\bibitem{tasinato} 
R.~Gregory, S.~L.~Parameswaran, G.~Tasinato and I.~Zavala,
  ``Lifshitz solutions in supergravity and string theory,''
   JHEP {\bf 1012} (2010) 047,
   [arXiv:1009.3445 [hep-th]].


\bibitem{hs10a} 
  H.~Singh,
  ``Holographic flows to IR Lifshitz spacetimes,''
  JHEP {\bf 1104}, 118 (2011)
  [arXiv:1011.6221 [hep-th]].

\bibitem{Bertoldi:2011zr} 
  G.~Bertoldi, B.~A.~Burrington, A.~W.~Peet and I.~G.~Zadeh,
  `Lifshitz-like black brane thermodynamics in higher dimensions,''
  Phys.\ Rev.\ D {\bf 83}, 126006 (2011)
  [arXiv:1101.1980 [hep-th]].

\bibitem{hs12}
  H.~Singh,
  ``Lifshitz/Schr\'odinger Dp-branes and dynamical exponents,''
  JHEP {\bf 1207}, 082 (2012)
  [arXiv:1202.6533 [hep-th]].




\bibitem{Narayan:2011az}
  K.~Narayan,
  ``Lifshitz-like systems and AdS null deformations,''
  Phys.\ Rev.\  D {\bf 84}, 086001 (2011)
  [arXiv:1103.1279 [hep-th]].

\bibitem{Cassani:2011sv} 
  D.~Cassani and A.~F.~Faedo,
  ``Constructing Lifshitz solutions from AdS,''
  JHEP {\bf 1105}, 013 (2011)
  [arXiv:1102.5344 [hep-th]].

\bibitem{Chemissany:2011mb}
  W.~Chemissany and J.~Hartong,
  ``From D3-Branes to Lifshitz Space-Times,''
  Class.\ Quant.\ Grav.\  {\bf 28}, 195011 (2011)
  [arXiv:1105.0612 [hep-th]].

\bibitem{Iizuka:2011hg}
  N.~Iizuka, N.~Kundu, P.~Narayan and S.~P.~Trivedi,
  ``Holographic Fermi and Non-Fermi Liquids with Transitions in Dilaton
  Gravity,''
  JHEP {\bf 1201}, 094 (2012)
  [arXiv:1105.1162 [hep-th]].


\bibitem{Song} W. Song and A. Strominger, ``Warped ADS3/Dipole-CFT Duality'',
arXiv:1109.0544 [hep-th].


\bibitem{Horowitz:2011gh}
  G.~T.~Horowitz and B.~Way,
  ``Lifshitz Singularities,''
  arXiv:1111.1243 [hep-th].


\bibitem{itzhaki} 
N. Itzhaki, J. Maldacena, J. Sonnenschein and S. Yankilowicz,
\prd{58}{1998}{046004}, \hepth{9802042}.

 

\bibitem{takaya11} 
  N.~Ogawa, T.~Takayanagi and T.~Ugajin,
  ``Holographic Fermi Surfaces and Entanglement Entropy,''
  JHEP {\bf 1201}, 125 (2012)
  [arXiv:1111.1023 [hep-th]].

\bibitem{subir11} 
  L.~Huijse, S.~Sachdev and B.~Swingle,
  ``Hidden Fermi surfaces in compressible states of gauge-gravity duality,''
  Phys.\ Rev.\ B {\bf 85}, 035121 (2012)
  [arXiv:1112.0573 [cond-mat.str-el]].

\bibitem{naray} K. Narayan, 
 `On Lifshitz scaling and hyperscaling violation in string theory',
arXiv:1202.5935 [hep-th].

\bibitem{kim}Bom Soo Kim,
`Schrödinger Holography with and without Hyperscaling Violation',
arXiv:1202.6062v1 [hep-th].

\bibitem{dey2012} 
  P.~Dey and S.~Roy,
  ``Lifshitz-like space-time from intersecting branes in string/M theory,''
  JHEP {\bf 1206}, 129 (2012)
  [arXiv:1203.5381 [hep-th]].

\bibitem{Lu:2012xu} 
  H.~Lu, Y.~Pang, C.~N.~Pope and J.~F.~Vazquez-Poritz,
  ``AdS and Lifshitz Black Holes in Conformal and Einstein-Weyl Gravities,''
  Phys.\ Rev.\ D {\bf 86}, 044011 (2012)
  [arXiv:1204.1062 [hep-th]].

\bibitem{Alishahiha:2012cm} 
  M.~Alishahiha and H.~Yavartanoo,
  ``On Holography with Hyperscaling Violation,''
  JHEP {\bf 1211}, 034 (2012)
  [arXiv:1208.6197 [hep-th]].

\bibitem{Alishahiha:2012qu}
  M.~Alishahiha, E.~O Colgain and H.~Yavartanoo,
 ``Charged Black Branes with Hyperscaling Violating Factor,''
  JHEP {\bf 1211} (2012) 137
  [arXiv:1209.3946 [hep-th]].

\bibitem{Gath:2012pg} 
  J.~Gath, J.~Hartong, R.~Monteiro and N.~A.~Obers,
  `Holographic Models for Theories with Hyperscaling Violation,''
  arXiv:1212.3263 [hep-th].

\bibitem{narayan12} 
  K.~Narayan, T.~Takayanagi and S.~P.~Trivedi,
  ``AdS plane waves and entanglement entropy,''
  JHEP {\bf 1304}, 051 (2013)
  [arXiv:1212.4328 [hep-th]];
  K.~Narayan,
  ``Non-conformal brane plane waves and entanglement entropy,''
  arXiv:1304.6697 [hep-th].

\bibitem{Edalati:2013tma} 
  M.~Edalati and J.~F.~Pedraza,
  ``Aspects of Current Correlators in Holographic Theories with Hyperscaling Violation,''
  arXiv:1307.0808 [hep-th].

\bibitem{Keeler:2013msa} 
  C.~Keeler, G.~Knodel and J.~T.~Liu,
  ``What do non-relativistic CFTs tell us about Lifshitz spacetimes?,''
  arXiv:1308.5689 [hep-th].

\bibitem{RT}
S. Ryu and T. Takayanagi, 
"Aspects of Holographic Entanglement Entropy",
JHEP 0608 (2006) 045, \hepth{0605073};
S. Ryu and T. Takayanagi, 
\prl{86}{2006}{181602}, \hepth{0603001}.

\bibitem{Maldacena:2013xja} 
  J.~Maldacena and L.~Susskind,
  ``Cool horizons for entangled black holes,''
  arXiv:1306.0533 [hep-th].

\bibitem{Faulkner:2013ana} 
  T.~Faulkner, A.~Lewkowycz and J.~Maldacena,
  ``Quantum corrections to holographic entanglement entropy,''
  arXiv:1307.2892 [hep-th].

\bibitem{hs13} H. Singh, "Lifshitz to AdS flow with interpolating $p$-brane solutions",
JHEP 1308 (2013) 097, \hepth{1305.3784}

\end{thebibliography}
\end{document}